# Magnetic field dependence of the critical current in stacked Josephson junctions. Evidence for fluxon modes in Bi$_2$Sr$_2$CaCu$_2$O$_{8+x}$ mesas.

V.M.Krasnov[a,b], N.Mros[a], A.Yurgens[a,c], and D.Winkler[a]

a) Department of Physics, Chalmers University of Technology, S-41296, Göteborg, Sweden,
b) Institute of Solid State Physics, 142432 Chernogolovka, Russia,
c) P.L.Kapitza Institute, Kosygina 2, 117334 Moscow, Russia

Modulation of the critical current across layers, $I_c(H)$, of stacked Josephson junctions (SJJs) as a function of an applied magnetic field parallel to the junction planes is studied theoretically and experimentally for different junction lengths and coupling parameters. It is shown that the $I_c(H)$ patterns of long SJJs are very complicated without periodicity in $H$. This is due to interaction between junctions in the stack. This, in turn, gives rise to the existence of multiple quasi-equilibrium Josephson fluxon modes and submodes which are different with respect to the symmetry of the phase and the fluxon sequence in SJJs. The critical current of long SJJs is multiple valued and is governed by switching between energetically close fluxon modes/submodes. Due to this, the probability distribution of the critical current may become wide and may consist of multiple maxima each representing a particular mode/submode. Experimentally, multiple branched $I_c(H)$ patterns and multiple maxima in the $I_c$ probability distribution were observed for Bi$_2$Sr$_2$CaCu$_2$O$_{8+x}$ intrinsic SJJs, which are in a good agreement with numerical simulations and support the idea of having different quasi-equilibrium fluxon modes/submodes in intrinsic SJJs.

## INTRODUCTION.

Properties of stacked Josephson junctions (SJJs) are of great interest both due to possible applications and from the general scientific point of view. A particular interest to SJJs was stimulated by the discovery of high-$T_c$ superconductors (HTSC). It is well known that the HTSC compounds have a layered structure with superconducting properties mainly confined to the Cu-O layers. Strong experimental evidence for that was provided by the observation of the intrinsic Josephson effect [1], which is attributed to Josephson coupling between atomic scale superconducting layers. However, it is far from being totally understood. For example, the superconducting 'gap' obtained from measurements of the intrinsic Josephson effect [1,2] is about two times smaller than that determined from ARPES [3] and tunnel measurements [4]. Furthermore, unambiguous observations of AC as well as DC Josephson effects remain to be obtained.

In this paper we study theoretically and experimentally the modulation of the critical current across layers, $I_c(H)$, of SJJs in a magnetic field parallel to the junction planes. Due to the fluxoid quantization the critical current of the Josephson junction (JJ) should exhibit the well known periodic 'Fraunhofer' modulation of $I_c(H)$ in parallel magnetic field, see e.g. Ref. [5], with periodicity

$$H_0 = \Phi_0 / L\Lambda^*, \qquad (1)$$

where $\Phi_0$ is a flux quantum, $L$ is the junction length, and $\Lambda^*$ is the effective magnetic thickness [6],

$$\Lambda_i^* = \Lambda_i - S_i - S_{i+1}, \qquad (2)$$

which for thin layered SJJs is equal to the interlayer distance, $s$. Here

$$\Lambda_i = t_i + \lambda_{si}\coth(d_i/\lambda_{si}) + \lambda_{si+1}\coth(d_{i+1}/\lambda_{si+1}), \qquad (3)$$

$$S_i = \lambda_{si}\operatorname{cosech}(d_i/\lambda_{si}). \qquad (4)$$

Here $\lambda_{si}$ and $d_i$ are the London penetration depth and the thickness of S-layers and $t_i$ is the tunnel barrier thickness. Hereafter, the subscript '$i$' of a quantity represents its number. Modulations of the $c$-axis $I_c(H)$ in HTSC intrinsic SJJs observed so far were not well defined [1,7,8]. Both periodic and aperiodic modulations of the $c$-axis resistivity $R(H)$ in YBa$_2$Cu$_3$O$_{7-x}$ were reported in Ref. [9]. However, they had a 'wrong' periodicity, and the authors had to assume different inter- and intra-unit-cell junctions with dimensions different from the crystal lattice space periodicity. The situation with low-$T_c$ SJJs is not much better. Very complicated Fraunhofer patterns were observed for Nb/AlO$_x$/Nb SJJs [7,10,11]. To our knowledge the only clear periodic modulation of the perpendicular resistivity $R(H)$ with periodicity $H_0 = \Phi_0/Ls$ was reported for Nb/Cu multilayers [12]. However, such a behavior was observed only in a small temperature range close to the critical temperature $T_c$. At lower temperatures the modulation in $R(H)$ becomes complicated without clear periodicity and with pronounced hysteresis [12].

Theoretically, the $I_c(H)$ dependence for 'small' SJJs, with $L$ much less than the Josephson penetration depth $\lambda_J$, was studied in Ref. [13]. It was shown that for small SJJs pure 'Fraunhofer' oscillations of $I_c(H)$ similar to that for a single JJ should be observed with a periodicity given by Eq.(1). In Ref. [14] the $I_c(H)$ of 'long' SJJs, $L \gg \lambda_J$, was studied. A monotonous power-law decrease of $I_c(H)$ was predicted. However, this assumes an absolutely rigid Josephson vortex lattice. More realistically, the critical current in long JJs is determined by fluxon entrance and their redistribution in the junction with increasing $H$, see e.g. [5]. In Ref. [12] it was argued that complicated aperiodic $I_c(H)$ patterns in long SJJs are caused by the asynchronous fluxon entrance in different junctions with the increase of $H$, which is a general property of long SJJs. In particular, fluxons in SJJs do not necessarily form a triangular lattice but rather may have more complicated arrangements as it can be seen from numerical simulations in Ref. [7]. Recently, it was shown that in SJJs there are multiple quasi-equilibrium fluxon configurations (modes) [6]. An evidence for the existence of such fluxon modes in $Bi_2Sr_2CaCu_2O_{8+x}$ mesas was obtained in Ref. [15]. In the dynamic state the existence of multiple fluxon modes in SJJs would result in multiple flux-flow branches in the c-axis I-V curves [6] and two-dimensional collective cavity resonances [16]. Multiple flux-flow branches have been observed for Nb/Cu multilayers [12] and $Bi_2Sr_2CaCu_2O_{8+x}$ mesas [17].

In the current paper we study fluxon modes in SJJs and their influence on the $I_c(H)$ patterns. We show that the $I_c(H)$ patterns in coupled long SJJs are very complicated and are determined by the process of switching between closely spaced fluxon modes, which occurs at the field intervals much less than $H_0$ given by Eq.(1) and without clear periodicity. In addition to the fluxon modes defined by the number of fluxons in each SJJ, we found that there are certain fluxon submodes, which are different with respect to the symmetry of the phase and the fluxon sequence. The $I_c(H)$ patterns for different junction lengths and coupling strengths are simulated numerically. The periodicity of $I_c(H)$ is restored when either $L$ becomes comparable to $\lambda_J$ or when coupling between junctions vanishes. Finally, we present the experimental c-axis $I_c(H)$ patterns for $Bi_2Sr_2CaCu_2O_{8+x}$ mesa structures. We observed multiple branches on the $I_c(H)$ patterns which are in good agreement with numerical simulations and support the idea of having multiple fluxon modes in intrinsic HTSC SJJs.

## THEORY

The behavior of SJJs can be described by the coupled sine-Gordon equation [18] which for a double SJJ in the static case can be written as

$$\lambda_{J1}^2 \varphi''_1 = sin(\varphi_1) - \frac{J_{c2} S_2}{J_{c1} \Lambda_1} sin(\varphi_2) - \frac{(\Lambda_1 - S_2)}{\Lambda_1} \frac{J_b}{J_{c1}}, \quad (5a)$$

$$\lambda_{J1}^2 \varphi''_2 = \frac{J_{c2} \Lambda_2}{J_{c1} \Lambda_1} sin(\varphi_2) - \frac{S_2}{\Lambda_1} sin(\varphi_1) - \frac{(\Lambda_2 - S_2)}{\Lambda_1} \frac{J_b}{J_{c1}}. \quad (5b)$$

Here $\varphi_{1,2}$ is the phase difference across junctions 1 and 2, respectively, the 'prime' denotes the spatial derivative, $J_{c1,2}$ and $J_b$ are the critical and the bias current densities, respectively and $\lambda_{J1}$ is the Josephson penetration depth of JJ1,

$$\lambda_{J1}^2 = \frac{\Phi_0 c}{8\pi^2 J_{c1} \Lambda_1}. \quad (6)$$

The coupling strength of double SJJs can be described by the dimensionless coupling parameter

$$S = \frac{S_2}{\sqrt{\Lambda_1 \Lambda_2}}. \quad (7)$$

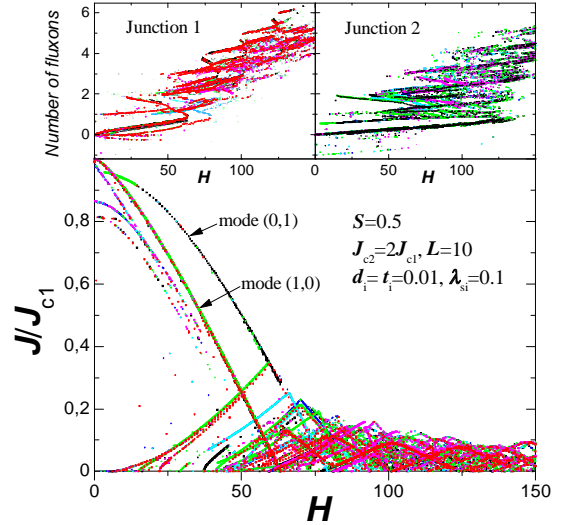

Fig. 1. Simulated $J_c(H)$ dependencies for a long strongly coupled double SJJs with nonidentical junctions $J_{c2}=2J_{c1}$. In the insets, the numbers of fluxons in junctions 1 and 2 are shown as a function of $H$. Multiple closely spaced fluxon submodes can be seen.

The thermodynamic equilibrium for a given $H$ is achieved at the minimum of Gibbs free energy, which for a double SJJ can be written as [6]

$$G(B) =$$
$$\sum_{i=1,2} J_{ci}(1 - cos(\varphi_i)) + \frac{1}{8\pi} \left[ B_{f1}^2 \Lambda_1 - 2 B_{f1} B_{f2} S_2 + B_{f2}^2 \Lambda_2 \right]$$
$$- \frac{H}{4\pi} \left[ B_{f1} \Lambda_1^* + B_{f2} \Lambda_2^* \right]. \quad (8)$$

Here $B_{f1,2}$ is the magnetic induction of the fluxon. From Eq. (8) it is seen that $G(B)$ is a bilinear form of $B_{f1,2}$ which can be minimized in different ways for different relations between $B_{f1}$ and $B_{f2}$. In other words, $G(B)$ has a particular minimum for each particular fluxon mode $(n_1,n_2)$, where $n_{1,2}$ are the number of fluxons in junctions 1 and 2, respectively.

Fig.1 shows simulated $J_c(H)$ dependencies for a double stack with an overlap geometry consisting of long, $L=10\lambda_{J1}$, nonidentical junctions $J_{c2}=2J_{c1}$ with thin S-layers and strong coupling $S\cong0.5$. The lengths are normalized to $\lambda_{J1}$, the current density is normalized to $J_{c1}$, and the magnetic field is normalized to $H_p=\Phi_0/(\pi\Lambda_1\lambda_{J1})$. In the insets we show the number of fluxons in junctions 1 and 2 determined via the total phase shift, $n_{1,2}=(\varphi_{1,2}(L)-\varphi_{1,2}(0))/2\pi$ at the maximum bias current. Simulations were made in the following way. The magnetic field was swept either from zero to the maximum field $H_{max}$ or in the opposite direction. For a particular $H$, a solution of Eq.(5) at zero bias current was found using a subsequent iteration, followed by the bias current increase. The critical current was determined as a maximum bias current value for which the numerical procedure converged. Fig.1 contains several runs with different initial conditions corresponding to different phase distributions in SJJs at $H=0$ and $H=H_{max}$. From

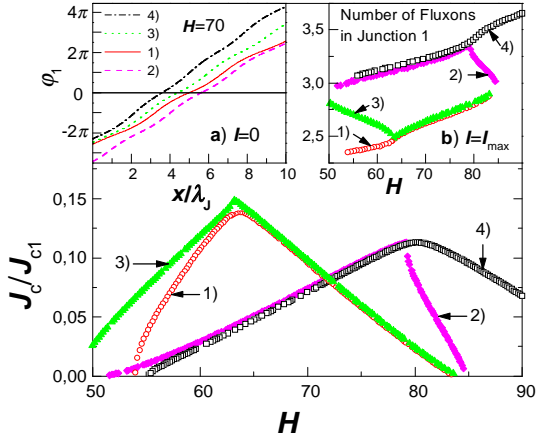
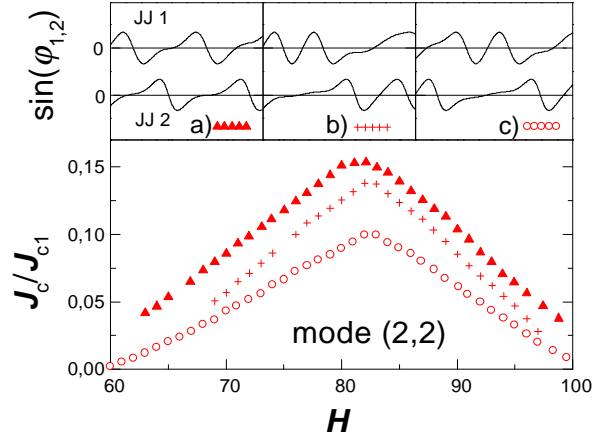

Fig.2. Four particular sub-branches of $J_c(H)$ corresponding to submodes of the (3,0) mode for the same SJJs as in Fig.1. In inset a) spatial phase distributions for the submodes are shown. Number of fluxons in JJ 1 is shown in inset b). The existence of several metastable solutions with different symmetry is seen.

Fig. 3. Three sub-branches $J_c(H)$ corresponding to submodes of the (2,2) mode with different fluxon sequence for the same SJJs as in Fig. 1. In the insets, current distributions in JJ 1 and JJ 2 for different submodes are shown.

Fig.1 it is seen that the $J_c(H)$ pattern is very complicated. It consists of a number of distinct and closely spaced branches. The insets in Fig.1 illustrate that those branches are characterized by certain fluxon configurations in the SJJs. Two particular $J_c(H)$ branches corresponding to a single fluxon entrance into JJ 1 and JJ 2, modes (1,0) and (0,1), respectively, are marked in Fig.1. From the insets in Fig.1 it is seen that there are closely spaced submodes even for the same number of fluxons in the junctions. From our analysis of spatial phase distributions we have found two main reasons for the existence of multiple submodes: (i) The sine-Gordon equation for long JJs allows several metastable solutions having different symmetry with respect to the junction center. (ii) The same number of fluxons in SJJs can be arranged in different sequences.

In Fig.2 we reproduce four particular sub-branches of $J_c(H)$ for the same SJJs as in Fig.1 for the (3,0) mode. The number of fluxons in JJ 1 is shown in inset b), from which it can be seen that each of the four $J_c(H)$ sub-branches corresponds to a particular (3,0) fluxon submode. In inset a) the spatial phase distributions for the four submodes are shown at $H=70$ at zero bias current. From Fig. 2 a) it is seen that in long SJJs there exist several solutions of the coupled sine-Gordon equation for the same $H$. The ambiguity in the solution of the sine-Gordon equation for long JJs was previously studied by Owen and Scalapino [19]. Recently, it was shown that several metastable solutions exist in a long JJ, which correspond to local minima of the free energy [20]. Solutions of this type are shown in Fig. 2 a). From the general properties of the sine-Gordon equation at zero bias current it can be shown that for a given solution there is a 'mirror' solution which is anti-symmetric with respect to the junction center. Such are solutions 2) and 3) in Fig. 2 a)

$$\varphi_{(2)}(x)=-\varphi_{(3)}(L-x). \qquad (9)$$

Solutions 1) and 4) do not have mirror solutions since
$$\varphi_{(1)}(x)=-\varphi_{(1)}(L-x),$$
$$\varphi_{(4)}(x)=-\varphi_{(4)}(L-x)+2\pi.$$

Application of the bias current removes the symmetry between solutions in Eq.(9) and results in each solution having its own critical current, as shown in Fig. 2.

The second reason for the existence of multiple submodes in long SJJs is inherent only to SJJs and is caused by fluxon interaction in different junctions. In Fig. 3, three sub-branches of $J_c(H)$ for the same SJJs as in Fig. 1 are shown corresponding to different submodes of the (2,2) mode. Current distributions in JJ 1 and JJ 2 are shown in the insets for $H=80$. From the insets it is seen that the submodes are different with respect to the fluxon sequence in SJJs. We note, that the self energies of the submodes shown in Fig. 3 are nearly equal, which means that SJJs can be in any of such states with nearly equal probability. The number of fluxon modes and submodes increases rapidly with increasing total number of junctions and fluxons. From mathematical statistics it follows that the number of fluxon modes (indistinguishable fluxon sequences) in a stack with $N$ non-identical junctions with $M$ fluxons in total is [6]

$$m=\frac{(N+M-1)!}{(N-1)!\,M!}, \qquad (10a)$$

and the number of submodes (distinguishable fluxon sequences) is
$$m_s=N^M. \qquad (10b)$$

The total number of fluxon submodes increases further due to the existence of several solutions with different symmetry, as discussed above, see Fig. 2, which are not included in Eq.(10). The number of submodes, and hence, the number of different $J_c(H)$ branches grows rapidly (exponentially) with $H$ due to the increase of $M$ in Eq.(10b). As a result the $J_c(H)$ patterns of long SJJs become very complicated and are determined by switching between closely spaced sub-branches corresponding to different submodes. Such switching occurs at field intervals much less than the period of the Fraunhofer pattern in a single JJ, Eq.(1). Fig.1 illustrates that the maximum of $J_c(H)$ in long SJJs decrease nearly monotonously with $H$ and do not have apparent periodicity.

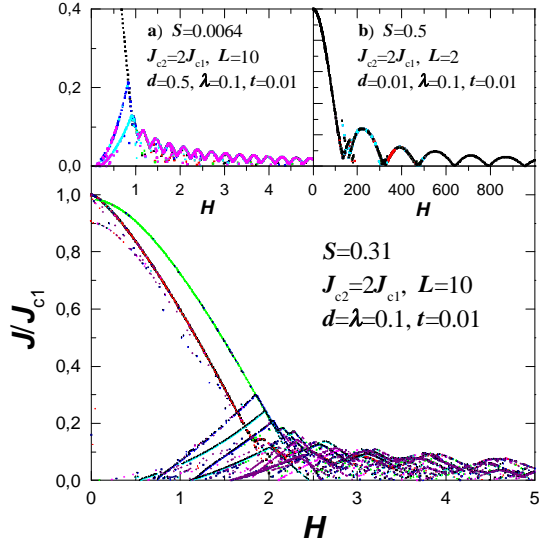
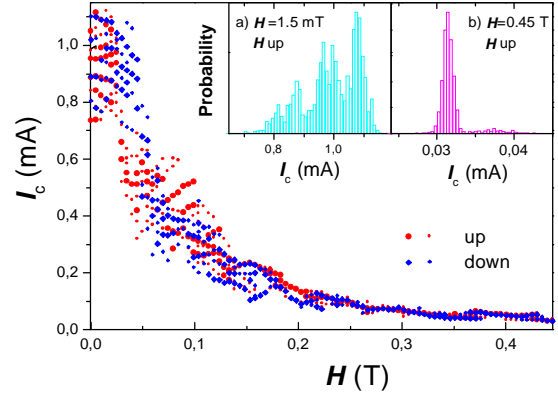

Fig.4. $J_c(H)$ pattern for a long double SJJs with intermediate coupling, $S \cong 0.31$. In inset a) $J_c(H)$ for a long double SJJs with small coupling is shown. In inset b) $J_c(H)$ of strongly coupled small SJJs is shown.

Fig.5. The $I_c(H)$ pattern for the 20 μm long $Bi_2Sr_2CaCu_2O_{8+x}$ mesa at $T$=4.2K. Circles correspond to increasing magnetic field and diamonds - to decreasing. Large symbols represent the main maxima while small symbols - the secondary peaks in the probability distributions. Multiple branches of $I_c(H)$ can be seen in Fig.5. In the insets the probability distributions of the critical current are shown for a) $H$ = 1.5 mT and b) $H$ = 0.45 T.

In Fig.4, the $J_c(H)$ pattern is shown for a long double SJJ with intermediate coupling, $S \cong 0.31$. It is seen that multiple fluxon submodes still exist. However, the difference between submodes is reduced with respect to the strong coupling case in Fig. 1 and some weak periodic modulation of $J_c(H)$ can be seen. With further decrease of coupling the difference between submodes disappears due to the vanishing fluxon interaction in neighboring SJJs. This is illustrated in inset a) of Fig. 4, where the $J_c(H)$ pattern for a long double SJJ with thick S-layers and small coupling is shown. In this case the $J_c(H)$ is simply defined by the minimum critical current of individual JJs in the stack and $J_c(H)$ has a periodic Fraunhofer modulation.

Periodic Fraunhofer patterns $J_c(H)$ can also be observed for small SJJs with $L$ comparable or less than $\lambda_J$ for arbitrary coupling strength [13]. In this case there are no fluxons and consequently no fluxon submodes that might complicate the $J_c(H)$ pattern. This is illustrated in the inset b) of Fig. 4, where the $J_c(H)$ pattern of strongly coupled small SJJs is shown. Apparent periodicity given by Eq.(1) is seen. A few sub-branches are still visible at low magnetic fields due to the finite length of the stack, $L$=2.

EXPERIMENT

For the study of intrinsic Josephson effect in HTSC, mesas with lengths 20 and 50 μm were fabricated on surfaces of $Bi_2Sr_2CaCu_2O_{8+x}$ single crystals [21]. The mesas typically contain 4-10 intrinsic SJJs. Several contacts on the top of the mesas allowed us to make four-probe measurements of I-V curves. The bias current was swept with a low frequency and the c-axis critical current, $I_c$, was measured with a 1 μV threshold criterion [22]. The magnetic field parallel to the intrinsic SJJs (ab plane) was generated by a superconducting coil. For each magnetic field value, 2048 switching events from zero voltage state were measured in series and were plotted as a probability distribution histogram, $P(I_c)$, defining the probability of having a particular critical current value. In the insets of Fig.5 the probability distributions of the critical current are shown for a) $H$ = 1.5 mT and b) $H$ = 0.45 T and at temperature $T$ = 4.2 K. It is seen, that at low magnetic fields the probability distribution is very wide (0.5 mA or more). Such a wide distribution cannot be simply explained by thermal fluctuations or external noise in a single JJ. The experimental setup has been tested by measuring low-$T_c$ Nb-AlO$_x$-Nb tunnel junctions, which show that external noise has a negligible effect. This can be seen also from the comparison with a high field case, see inset b) in Fig.5, for which the width of the probability distribution is much smaller (about 2.5μA). Note, that the noise level does not decrease with increasing field. From inset a) in Fig.5 it is seen that the critical current probability distribution consists of multiple superimposed maxima (a certain number of switching events with larger $I_c$ can be also seen in the inset b)). Such a behavior is consistent with the idea of the existence of multiple quasi-equilibrium fluxon modes in intrinsic SJJs with different critical currents. In this case each maximum corresponds to a particular fluxon mode. The state of SJJs is not uniquely defined by external conditions, but rather, it can be described only statistically with a certain probability of being in any of the quasi-equilibrium fluxon modes, defined by the energy of the mode. In Fig. 5, we plotted the values of c-axis $I_c$ corresponding to maxima in the probability distribution versus the applied magnetic field for the 20 µm long mesa. The length of the mesa is much larger than the Josephson penetration depth which is estimated to be less than 1 µm. Circles correspond to increasing magnetic field and diamonds - to decreasing. Large symbols represent the main maxima while the small symbols - the secondary peaks in the probability distributions. From Fig. 5 it is seen that the $I_c(H)$ pattern of $Bi_2Sr_2CaCu_2O_{8+x}$ mesa is complicated and does not exhibit clear periodicity. However, several distinct branches of $I_c(H)$ can be seen in Fig.5. Similar behavior of $I_c(H)$ has been observed for all studied mesas. The $I_c(H)$ patterns exhibit a certain hysteresis and a prehistory dependence and depend on the sample geometry. All this is in a good qualitiative agreement with our numerical simulation and supports the idea of existence of multiple quasi-equilibrium fluxon modes in HTSC intrinsic SJJs.

In conclusion, we have shown that the $I_c(H)$ patterns of $Bi_2Sr_2CaCu_2O_{8+x}$ mesas are consistent with the existence of Josephson effect between atomic scale superconducting layers so that the single crystal behaves as intrinsic SJJs. On the other hand, when the length of SJJs is larger than the Josephson penetration depth the behavior of $I_c(H)$ is much more complicated than that of a single JJs and does not exhibit the periodicity in $H$. This is due to the existence of multiple quasi-equilibrium fluxon modes with different number of fluxons in SJJs. From the theoretical analysis we have found that there exist various fluxon submodes different with respect to the symmetry of the phase and the fluxon sequence. The critical current of long SJJs is multiple valued and is governed by switching between quasi-equilibrium fluxon modes and submodes. Due to the fact that energetically close modes/submodes can have different $I_c$, the probability distribution of the critical current becomes very wide and consists of multiple maxima, each representing a particular mode/submode. This is exactly what we have observed experimentally for long $Bi_2Sr_2CaCu_2O_{8+x}$ mesas.

## ACKNOWLEDGMENTS

The work was supported by the Swedish Superconductivity Consortium and in part by the Russian Foundation for Basic Research under the Grant No. 96-02-19319.